\documentclass[twocolumn,showpacs,floatfix]{revtex4}%
\usepackage{graphicx}%
\usepackage{amsmath}%
\usepackage{amsfonts}%
\usepackage{amssymb}
\usepackage{bm}

\def\q{{ {\bm q} }}

\def\w{{\omega}}
\def\a{{\alpha}}

\allowdisplaybreaks[4]

\begin{document}
\title{
$S$-wave Superconductivity due to Orbital and Spin fluctuations
in Fe-pnictides: 
Self-Consistent Vertex Correction with Self-Energy (SC-VC$_\Sigma$) Analysis 
}
\author{
Seiichiro \textsc{Onari}$^{1}$, Hiroshi \textsc{Kontani}$^{2}$,
Sergey V. Borisenko$^{3}$, Volodymyr B. Zabolotnyy$^{3}$ and 
Bernd B{\"u}chner$^{3}$
}

\date{\today }

\begin{abstract}
To understand the amazing variety of the superconducting states
of Fe-based superconductors,
we analyze the multiorbital Hubbard models for LaFeAsO and LiFeAs
going beyond the random-phase approximation (RPA),
by calculating the vertex correction (VC) and self-energy correction.
Due to the spin+orbital mode coupling described by the VC,
both orbital and spin fluctuations mutually develop,
consistently with the experimental phase diagram 
with the orbital and magnetic orders.
Due to both fluctuations, the $s$-wave gap function with sign-reversal 
($s_{\pm}$-wave), without sign-reversal ($s_{++}$-wave),
and nodal $s$-wave states are obtained, compatible with the
experimental wide variety of the gap structure.
Thus, the present theory provides a microscopic
derivation of the normal and superconducting phase diagram 
based on the realistic Hubbard model.

\end{abstract}

\address{
$^1$ Department of Applied Physics, Nagoya University,
Furo-cho, Nagoya 464-8603, Japan. 
\\
$^2$ Department of Physics, Nagoya University,
Furo-cho, Nagoya 464-8602, Japan. 
\\
$^3$ Leibniz-Institute for Solid State Research, IFW-Dresden, 
D-01171 Dresden, Germany 
}
 

\sloppy

\maketitle

One of the main characteristic feature of 
Fe-based superconductors is their
amazing variety of the superconducting gap structure.
For example, isotropic fully-gapped $s$-wave state is 
realized in optimally-doped Co-doped and K-doped BaFe$_2$As$_2$
\cite{Hashimoto,Tohoku-ARPES,Shimo-Science},
while nodal $s$-wave states are expected 
in under and over-doped compounds
\cite{Taillefer}.
In optimally P-doped  BaFe$_2$As$_2$, in contrast,
nodal $s$-wave state is observed by angle-resolved thermal conductivity 
\cite{Kasahara-angle}
and angle-resolved photoemission spectroscopy (ARPES)
\cite{Shimojima,Yoshida,Feng}.
Such variety of the gap structure would indicates the 
presence of competing pairing interactions in Fe-based superconductors
\cite{Saito-3Dgap,Hirschfeld2,Chubukov2}.

To clarify the pairing mechanism in strongly correlated systems,
its normal state phase diagram should be understood.
The coincidence of the structure (or orbital) and 
magnetic quantum critical points (QCPs) 
indicates the coexistence of orbital and spin fluctuations
in Fe-based superconductors.
The large softening of the shear modulus $C_{66}$ 
above the structure transition temperature $T_S$
is a direct evidence of orbital fluctuations
\cite{Fernandes,Kontani-rev,Yoshizawa,Goto,KFe2As2,Raman},
and the orbital order or polarization had been observed 
in various compounds by polarized ARPES measurements
\cite{Shen,Shimo-arXiv}.
As for the superconductivity, the $s$-wave state with (without) sign reversal
is caused by spin (orbital) fluctuations
\cite{Kuroki,Mazin,Hirschfeld,Chubukov,Kontani-RPA,Saito-RPA,Onari-VC}.
In BaFe$_2$(As,P)$_2$, the absence of the horizontal node
on the hole-doped Fermi surface (h-FS) with $3z^2-r^2$-orbital character
\cite{Shimojima,Yoshida}
indicated the importance of the orbital fluctuations 
\cite{Saito-3Dgap}, and 
the loop-nodes on the electron-type Fermi surfaces (e-FSs)
\cite{Kasahara-angle,Shimojima,Yoshida}
can be explained by the competition
between different spin fluctuations \cite{Hirschfeld2,Chubukov2} 
or that between spin and orbital fluctuations
\cite{Saito-3Dgap}.

The normal state phase diagram with
{\it non-magnetic} structure transition cannot be
explained by the random-phase-approximation (RPA)
based on the multiorbital Hubbard model.
However, it can be explained
by considering the vertex correction (VC), since
the spin fluctuations induce the charge quadrupole order
($O_{x^2-y^2}\equiv n_{xz}-n_{yz}\ne0$) 
owing to the spin+orbital mode coupling described by the VC.
This mechanism has been demonstrated by the diagrammatic
\cite{Onari-VC,Ohno-SCVC}
as well as the renormalization group \cite{Tsuchiizu} methods
in several multiorbital models.
Similar spin-orbital coupling beyond the RPA
had also been discussed in Refs. \cite{Dagotto,KK}.
We stress that spin fluctuations could also induce the spin quadrupole 
($\phi={\bm s}_{\rm A}\cdot {\bm s}_{\rm B}$) if $J_1\sim2J_2$ 
is realized \cite{Fernandes},
so the correct quadrupole order should be elucidated experimentally.
The strain-quadrupole order coupling 
derived from the fitting of $C_{66}$ is very large, which would
be natural for the charge quadrupole scenario \cite{Kontani-rev}.
Now, the next significant challenge is to elucidate the
pairing mechanism of the Fe-based superconductors 
by taking the VC into account.

In this paper, we develop a theory beyond the RPA,
by calculating both the VC and the self-energy $\Sigma$ self-consistently.
By this ``self-consistent VC+$\Sigma$ (SC-VC$_\Sigma$) method'',
we obtain the mutual development of spin and orbital fluctuations.
Therefore, both the $s$-wave state with sign change
($s_{\pm}$-wave) and that without sign change ($s_{++}$-wave),
both of which are promising canditate pairing states,
are naturally reproduced based on the two different realistic Hubbard models.
No additional interactions such as the quadrupole interaction 
\cite{Kontani-RPA} were introduced to the models.
The obtained smooth $s_{++}\leftrightarrow s_\pm$ crossover 
could explain the wide variety of gap structures 
in Fe-based superconductors.




In the following, we explain the SC-VC$_\Sigma$ method.
First, we employ the five-orbital Hubbard model for LaFeAsO
introduced in Ref. \cite{Kuroki}.
We denote $d$-orbitals $3z^2-r^2$, $xz$, $yz$, $xy$, and $x^2-y^2$
as 1, 2, 3, 4 and 5, respectively. Hereafter, $x,y$-axes are along the nearest Fe-Fe direction.
The Fermi surfaces are mainly composed of orbitals 2, 3 and 4 
\cite{Kontani-Soft}.
The susceptibility for the charge (spin)
channel is given by the following $25\times25$ matrix form 
in the orbital basis:
\begin{eqnarray}
{\hat \chi}^{c(s)}(q)={\hat {\chi}}^{{\rm irr},c(s)}(q)
 (1-{\hat \Gamma}^{c(s)}{\hat {\chi}}^{{\rm irr},c(s)}(q))^{-1} ,
\label{eqn:chi}
\end{eqnarray}
where $q=(\q,\w_l=2\pi l T)$, and
${\hat \Gamma}^{c(s)}$ represents the Coulomb interaction
for the charge (spin) channel composed of $U$, $U'$ and $J$ 
given in Refs. \cite{Kontani-RPA,Saito-RPA,Takimoto}.
The irreducible susceptibility in Eq. (\ref{eqn:chi}) is given as
\begin{eqnarray}
{\hat {\chi}}^{{\rm irr},c(s)}(q)= {\hat \chi}^{0}(q)+{\hat X}^{c(s)}(q) ,
\label{eqn:irr}
\end{eqnarray}
where $\chi^{0}_{ll',mm'}(q)=-T\sum_p G_{lm}(p+q)G_{m'l'}(p)$ is the bare bubble,
and the second term
is the VC, which is essential to produce the orbital fluctuations as
discussed in Ref. \cite{Onari-VC}. 
In the SC-VC$_\Sigma$
method, Green's function $\hat{G}$ is given by Dyson's equation
$\hat{G}=(\hat{G}_0^{-1}-\hat{\Sigma})^{-1}$, where $\hat{G}_0$ is the
bare Green's function and $\hat{\Sigma}$ is the self-energy.
(In the SC-VC method, we put $\hat{\Sigma}=0$.)
In order to measure the distance from the criticality,
we introduce the charge (spin) Stoner factor $\a^{c(s)}_\q$, which is 
the largest eigenvalue of 
${\hat \Gamma}^{c(s)}{\hat {\chi}}^{{\rm irr},c(s)}(\q)$ at $\w_l=0$ 
\cite{Kontani-RPA}:
The charge (spin) susceptibility diverges when
$\a^{c(s)}\equiv {\rm max}_\q\{\a^{c(s)}_\q\}=1$.

Here, we introduce VCs
due to the Aslamazov-Larkin (AL) terms,
which is the second order term with respect to ${\hat \chi}^{c,s}$
and becomes important near the QCP
\cite{Moriya,Onari-VC}.
The VCs for charge(spin) sector are denoted as ${\hat X}^{c(s)}(q)\equiv {\hat X}^{\uparrow,\uparrow}(q) +(-)
{\hat X}^{\uparrow,\downarrow}(q)$.
The AL term for the charge sector,
$X_{ll',mm'}^{{\rm AL},c}(q)$, is given as
\begin{eqnarray}
& &\frac{T}2\sum_{k}\sum_{a\sim h}
\Lambda_{ll',ab,ef}(q;k)\{  {V}_{ab,cd}^c(k+q){V}_{ef,gh}^c(-k)
\nonumber \\
& &\ \ +3{V}_{ab,cd}^s(k+q){V}_{ef,gh}^s(-k) \}
\Lambda_{mm',cd,gh}'(q;k) ,
 \label{eqn:ALexample}
\end{eqnarray}
where 
${\hat V}^{s,c}(q)\equiv{\hat \Gamma}^{s,c}
+ {\hat\Gamma}^{s,c}{\hat\chi}^{s,c}(q){\hat\Gamma}^{s,c} $, 
${\hat \Lambda}(q;k)$ and ${\hat \Lambda}'(q;k)$ are the three-point vertex
made of three Green functions\cite{Onari-VC}.
We include all $U^2$-terms without the double counting
to obtain reliable results. 
Note that we neglect $\hat{X}^{{\rm AL},s}$ because the
contribution of $\hat{X}^{{\rm AL},s}$ is much smaller than that of 
$\hat{X}^{{\rm AL},c}$ \cite{Onari-VC,Ohno-SCVC,Tsuchiizu}.

The $5\times 5$ self-energy matrix $\hat{\Sigma}$
in the fluctuation-exchange (FLEX) approximation is given by
\begin{eqnarray}
\Sigma_{lm}(k)=T\sum_{q}\sum_{l',m'}V^\Sigma_{ll',mm'}(q)G_{l'm'}(k-q),
\end{eqnarray}
where $\hat{V}^\Sigma(q)$ is the effective interaction for the self-energy:
$\displaystyle
\hat{V}^\Sigma(q)=\frac{3}{2}\hat{V}^s(q)+\frac{1}{2}\hat{V}^c(q)
-\frac{1}{4}(\hat{\Gamma}^c-\hat{\Gamma}^s)\hat{\chi}^0(q)(\hat{\Gamma}^c-\hat{\Gamma}^s)-\hat{\Gamma}^b\hat{\chi}^0(q)\hat{\Gamma}^b$,
where $\Gamma^b_{ll',mm'}=-U'+J$ for $l=l'\ne m=m'$ and
$\Gamma^b_{ll',mm'}=0$ for others.
The third and fourth terms of right hand side in $\hat{V}^\Sigma(q)$ are
required to cancel the double counting in the 2nd order diagrams.
By solving above equations,
we obtain the susceptibilities and self-energy self-consistently.

In our calculation, we neglect the Maki-Thompson (MT) terms since 
it is much smaller than the AL term as explained in Ref. \cite{Onari-VC}.
The dominance of the AL term is also verified by 
recent renormalization group study \cite{Tsuchiizu}. 
We use $\hat{G}_0$ 
in calculating $\Lambda$ and $\Lambda'$ in Eq. (\ref{eqn:ALexample}) 
since they are underestimated at high temperatures ($T\sim0.05$)
due to large quasiparticle damping
Im$\Sigma(\q,-i\delta) \propto T$.

\begin{figure}[!htb]
\includegraphics[width=.99\linewidth]{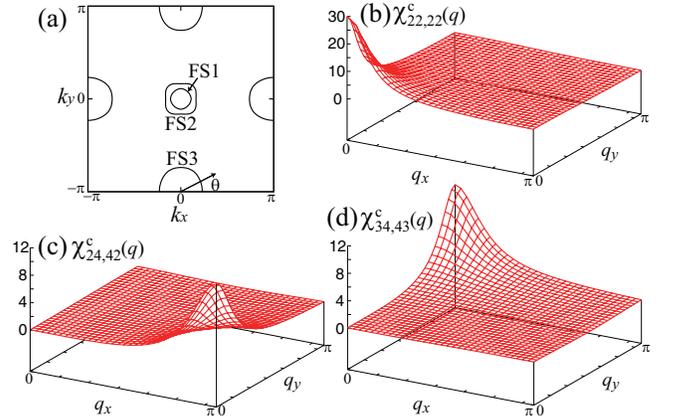}
\caption{(color online)
(a)FSs of the LaFeAsO five-orbital model for $n=6.1$, where $\theta$ is denoted by
 the azimuthal angle from $k_x$ axis.
(b)$\chi^c_{22,22}(\q)$, (c)$\chi^c_{24,42}(\q)$ and (d)$\chi^c_{3443}(\q)$ given by the SC-VC$_\Sigma$
 method for $J/U=0.15$, $\a^s=0.96$, and $\a^c=0.97$.
}
\label{fig:fig1}
\end{figure}

We use $64\times64$ $\bm k$-meshes and $256$ Matsubara frequencies
at $T=0.05$ eV, and set the unit of energy as eV.
Hereafter, we put the constraint $U=U'+2J$.
The Fermi surfaces in the LaFeAsO model for $n=6.1$ are shown in
Fig. \ref{fig:fig1} (a), where $\theta$ is denoted by
 the azimuthal angle from $k_x$ axis.
Figure \ref{fig:fig1} (b), (c) and (d) shows the 
obtained $\chi_{22,22}^c(\q)$, $\chi_{24,42}^c(\q)$ and
$\chi_{34,43}^c(\q)$
for $J/U=0.15$ and $U=2.2$ ($\a_s=0.96$ and $\a_c=0.97$)
using the SC-VC$_\Sigma$ method, respectively.
The development of the ferro-orbital fluctuations 
explains the softening of the shear modulus $C_{66}$ and structure transition,
and both ferro- and antiferro (AF)-orbital fluctuations are the driving force
of the $s_{++}$-wave state.
Here, orbital fluctuations can develop
for much large $J/U$ compared to the SC-VC method \cite{Onari-VC},
since the value of $U$ for the ordered state increases due to the self-energy,
and therefore ${\hat X}^{{\rm AL},c}\propto U^4$ is enlarged.
We verified that similar results are obtained for $6.0\le n\le6.1$,
even if the h-FS at $(\pm\pi,\pm\pi)$ appears.


Next, we focus on the superconducting gap function.
The linearized Eliashberg equation in the absence of impurities is given by
\begin{eqnarray}
\lambda_{\rm E}\phi_{ll'}(k)&=-&T\sum_{k',m_i}V^{\rm E}_{lm_1,m_4l'}(k-k')
G_{m_1m_2}(k')\nonumber\\
& &\times\phi_{m_2m_3}(k')G_{m_4m_3}(-k')
\label{eqn:Eliash}
\end{eqnarray}
where $\phi_{ll'}(k)$ is the anomalous self-energy and
$\lambda_{\rm E}$ is the eigenvalue that reaches unity at
$T=T_c$. When $T$ is fixed, $\lambda_{\rm E}$ is considered to be guide
of stability of the superconductivity, {\it
i.e.,} the larger eigenvalue $\lambda_{\rm E}$
corresponds to the higher $T_c$.
The pairing interaction ${\hat V}^{\rm E}$ in Eq. (\ref{eqn:Eliash}) is
\begin{equation}
\hat{V}^{\rm E}(q)=\frac{3}{2}\hat{\Gamma}^s\hat{\chi}^s(q)\hat{\Gamma}^s-\frac{1}{2}\hat{\Gamma}^c\hat{\chi}^c(q)\hat{\Gamma}^c+\hat{V}^{(1)},
\label{eqn:W}
\end{equation}
where $\hat{\chi}^{s,c}$ is given by the SC-VC$_\Sigma$ method for $n_{\rm imp}=0$, 
 and $\hat{V}^{(1)}$ denotes the first order term.
The first (second) term in Eq. (\ref{eqn:W}) works to 
set $\Delta_{\rm FS1,2}\cdot \Delta_{\rm FS3,4}<0$ ($>0$)
\cite{Takimoto}.

Finally, we obtain a gap function
$\hat{\Delta}(\bm{k})=\hat{\phi}(\bm{k},\omega_n=\pi T)/\hat{Z}(\bm{k})$
in the band representation,
where mass enhancement factor $\hat{Z}$ is given as
$Z_{\alpha}(\bm{k})=1-\frac{{\rm Im}\Sigma_{\alpha}(\bm{k},\omega_n=\pi
T)}{\pi T}$ in a band $\alpha$. We obtain $Z\sim 3$ in the present study.
We note that the absolute value of $\Delta$ is not important since the
Eliashberg Eq. (\ref{eqn:Eliash}) is linearized.


In Eq. (\ref{eqn:W}), $\hat{V}^{(1)}=\frac{1}{2}\hat{\Gamma}^s-\frac{1}{2}\hat{\Gamma}^c$
represents the first-order terms with respect to the Coulomb interaction.
This term gives the Anderson-Morel pseudopotential
$\mu^*\approx UN(0)[1+UN(0)\ln(W_{\rm band}/\w_{0})]^{-1}$,
where $N(0)$ is the density-of-states (DOS) at the Fermi level,
$W_{\rm band}$ is the bandwidth, and $\w_{0}$ is the 
energy-scale of the orbital and spin fluctuations
\cite{Anderson-Morel}:
$\w_{0}\sim T$ is expected in optimally-doped compounds 
close to the orbital and magnetic QCPs.
Although $\mu^*$ suppresses the $s_{++}$-wave state,
we can expect that this term is approximately canceled out by 
the weak e-ph interaction $\lambda_{\rm e-ph}\ (\sim0.2)$.

\begin{figure}[!htb]
\includegraphics[width=.99\linewidth]{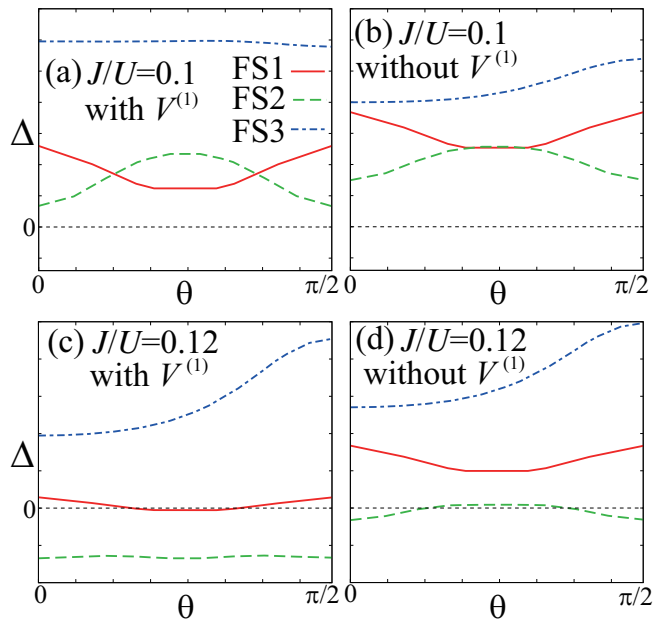}
\caption{(color online)
(a) $\theta$ dependences of gap functions for $J/U=0.1$  with
 $\hat{V}^{(1)}$ and (b) those without $\hat{V}^{(1)}$.
(c) $\theta$ dependences of gap functions for $J/U=0.12$ with
 $\hat{V}^{(1)}$ and (d) those without $\hat{V}^{(1)}$.
}
\label{fig:SCgap}
\end{figure}

Hereafter, we calculate the superconducting gap functions 
based on the gap equation in Eq. \ref{eqn:W},
by fixing the ratio $J/U$ while choosing $U$ so as to 
satisfy $\a_c=0.97$.
Figure \ref{fig:SCgap} (a) and (b) show the 
obtained gap structures for $J/U=0.1$ ($U=2.0$; $\a_s=0.91$)
with $\hat{V}^{(1)}$ ($\lambda_{\rm E}=0.35$)
and those without $\hat{V}^{(1)}$ ($\lambda_{\rm E}=0.51$), respectively. 
The dropping of $\hat{V}^{(1)}$ will be justified since we expect
that $\mu^*$ is as small as $\lambda_{\rm e-ph}\ (\sim0.2)$ 
due to the retardation.
In both cases, fully-gapped $s_{++}$-wave states are obtained.
In (b), gap functions of FS1 and FS2 are as large as that of FS3
since $\hat{V}^{(1)}$ is repulsive interaction and unfavorable to the
$s_{++}$-wave state.
The $s_{++}$-wave state is derived from the strong developments of 
$\chi_{22,22}(q)$ and $\chi_{24,42}(q)$ shown in Fig.\ref{fig:fig1}. 

Next, we move to $J/U=0.12$, where the spin fluctuations are stronger than that
 for $J/U=0.1$. Figure \ref{fig:SCgap} (c) and (d) show the 
obtained gap structures for $J/U=0.12$ ($U=2.1$; $\a_s=0.93$)
 with $\hat{V}^{(1)}$ ($\lambda_{\rm E}=0.49$) and those without
$\hat{V}^{(1)}$ ($\lambda_{\rm E}=0.57$), respectively. 
In both (c) and (d), the obtained states are the nodal $s$-wave states
which is intermediate between $s_{++}$-wave and $s_\pm$-wave states
due to the competition between orbital and spin fluctuations.
We note that the relation $J/U=0.12-0.15$ is
estimated by the first principle calculation \cite{Miyake}.

\begin{figure}[!htb]
\includegraphics[width=.99\linewidth]{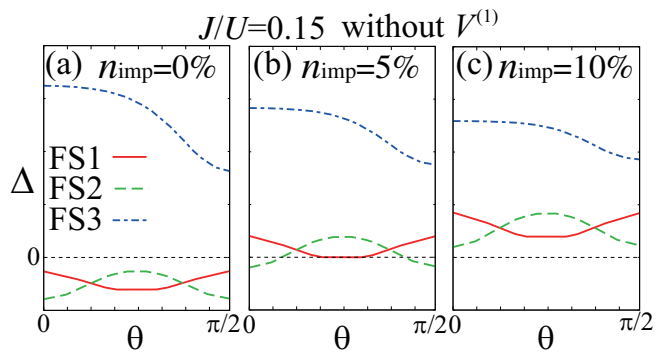}
\caption{(color online)
(a) $\theta$ dependences of gap functions without $\hat{V}^{(1)}$ for
 $J/U=0.15$ with $n_{\rm imp}=0$\%, (b) those with $n_{\rm imp}=5$\%, and
 (c) those with $n_{\rm imp}=10$\%.
}
\label{fig:SCgap-imp}
\end{figure}

We also study the impurity effect on the superconducting state,
by introducing the impurity $T$-matrix into Eq. (\ref{eqn:Eliash})
according to Refs. \cite{Onari-impurity,Kontani-RPA,Yamakawa-imp}.
Here, we discuss the impurity-induced 
$s_{\pm} \rightarrow s_{++}$ crossover for $J/U=0.15$
($U=2.2$; $\a_s=0.96$; $\lambda_{\rm E}=0.33$), in which 
the spin and orbital fluctuations are comparable.
In the absence of impurities ($n_{\rm imp}=0$),
the $s_{\pm}$-wave state is realized as shown in Fig. \ref{fig:SCgap-imp} (a).
As increasing $n_{\rm imp}$, 
the crossover between $s_{++}$-wave and $s_{\pm}$-wave states is expected,
as discussed in Refs. \cite{Kontani-RPA,Saito-3Dgap}.
In fact, Fig. \ref{fig:SCgap-imp} (b) and (c)  shows the
obtained gap functions for $n_{\rm imp}=5$\% and $10$\%, respectively,
when the impurity potential is $I=1$eV.
Since the obtained $\lambda_{\rm E} \ (\sim 0.33)$ 
is almost unchanged for $n_{\rm imp}=0\sim$10\%,
the impurity-induced $s_{\pm} \rightarrow s_{++}$ crossover 
will be realized with small change in $T_{\rm c}$.
Since the impurity concentration to realize the crossover is scaled with $T$,
$n_{\rm imp}=10$\% for $T=0.05$ would correspond to $n_{\rm imp}\sim1$\% 
for $T=T_c\sim 0.005$. 
Therefore, the $s_{++}$-wave state will be realized 
in various iron-based superconductors with finite randomness.

\begin{figure}[!htb]
\includegraphics[width=.99\linewidth]{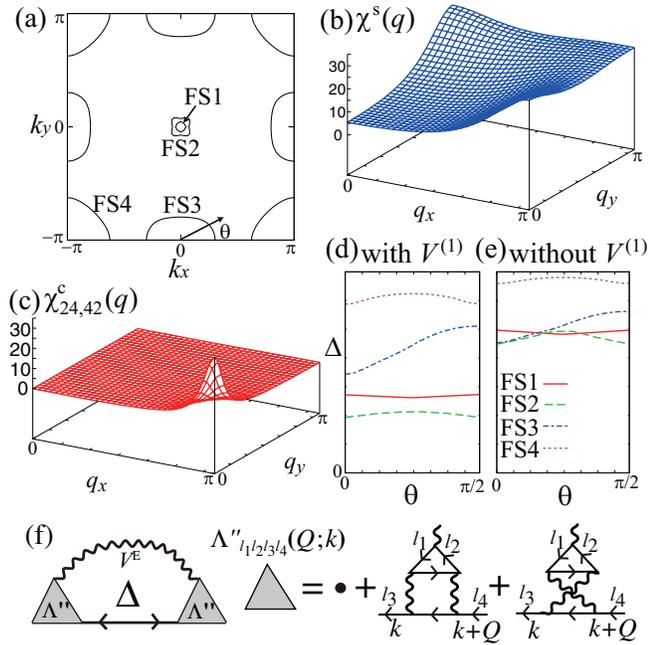}
\caption{(color online)
(a)FSs of the LiFeAs model for $n=6.0$. 
(b)$\chi^s(\q)=\sum_{l,m}\chi^s_{ll,mm}(\q)$, 
(c)$\chi^c_{24,42}(\q)$ given by SC-VC$_\Sigma$ method 
in the LiFeAs model for $J/U=0.12$, $\a^s=0.92$, and 
$\a^c=0.97$ at $T=0.03$.
(d) $\theta$ dependences of gap functions for
 $J/U=0.12$, $n_{\rm imp}=0$\% at $T=0.03$ with $\hat{V}^{(1)}$, and (e)
 those without $\hat{V}^{(1)}$.
(f) Feynman diagram of $\hat{\Delta}$ and VC
 $\hat{\Lambda}''$ for the charge sector $\hat{V}^c$ in
 the pairing interaction $\hat{V}^{\rm E}$.
}
\label{fig:Li111}
\end{figure}

In the following, 
we study the superconducting state of LiFeAs,
using the tight-binding model given by fitted to the ARPES data 
of LiFeAs \cite{Borisenko}.  
The FSs in the model are shown as Fig. \ref{fig:Li111} (a) for $n=6.0$, 
in which the large $d_{xy}$-orbital h-FS4
around $\bm{k}=(\pi,\pi)$ is added to the model of LaFeAsO. 
Within the RPA, the $s_{\pm}$-wave state is
favored by the h-FS4 \cite{Kuroki2}.
However, as shown in Fig. \ref{fig:Li111} (b) and (c) for $J/U=0.12$,
$U=1.2$ at $T=0.03$ ($\a_s=0.92$ and $\a_c=0.97$), 
the AF-spin and AF-orbtital fluctuations are strongly enhanced in the
SC-VC$_\Sigma$ method.
In contrast, the ferro-orbital fluctuations 
are relatively small, consistently with the absence of 
the structure transition in LiFeAs.
The gaps obtained for $J/U=0.12$, $n_{\rm imp}=0$\% at $T=0.03$
are $s_{++}$-wave mediated by the AF-orbital fluctuation
as shown in Fig. \ref{fig:Li111} (d) with $\hat{V}^{(1)}$ 
($\lambda_{\rm E}=0.40$) and (e) without $\hat{V}^{(1)}$ ($\lambda_{\rm E}=0.53$). 
Thus, the fully-gapped $s_{++}$-wave state is realized for $J/U=0.12$
even for $n_{\rm imp}=0$.

Here, the calculation temperature is much higher than $T_c$
since the SC-VC$_\Sigma$ method is heavy numerical calculation.
For this reason, the obtained gap is nearly isotropic on each FS.
It is our important future problem 
to study the experimental large gap anisotropy 
at much lower temperatures.

In the present study,
we dropped the VC for the electron-boson coupling constant in the gap equation, 
shown by $\hat{\Lambda}''(q;k)$ in Fig. \ref{fig:Li111} (f).
Recently, we had verified that $|\hat{\Lambda}''(q;k)|$ 
due to the AL type diagram in Fig. \ref{fig:Li111} (f) 
is much larger than unity for the charge sector \cite{SDW+SC}.
This fact means the violation of the Migdal's theorem.
Then, $\lambda_{\rm E}$ for the $s_{++}$-wave state is enlarged
since the charge pairing interaction 
$\frac{1}{2}\hat{\Gamma}^c\hat{\chi}^c(q)\hat{\Gamma}^c$,
given by the second term in Eq. (\ref{eqn:W}),
is multiplied by $|\hat{\Lambda}''(q;k)|^2$.
Therefore, the $s_{++}$-wave state is further stabilized by 
the VC for the electron-orbiton coupling, going beyond the Eliashberg theory.

In summary,
we have studied the normal and superconducting states of LaFeAsO and LiFeAs
using the SC-VC$_\Sigma$ theory.
We obtain both the $s_{++}$- and $s_\pm$-wave gap,
both of which are promising candidate pairing states,
based on two very different but realistic tight-binding Hubbard models
with $J/U\lesssim0.15$.
No additional interactions such as the quadrupole interaction
were introduced.
We stress that a smooth  $s_{++}\leftrightarrow s_\pm$ crossover could be
realized by introducing the impurities or magnetic ordered state,
consistently with the robustness of $T_{\rm c}$ 
\cite{Sato-imp,irradiation,Nakajima,FCZhang}
and the wide variety of gap structures in Fe-based superconductors.


\acknowledgements
We are grateful to 
A. Chubukov, P.J. Hirschfeld, J. Schmalian and R. Fernandes
for useful discussions.
This study has been supported by Grants-in-Aid for Scientific 
Research from MEXT of Japan.
Part of numerical calculations were
performed on the Yukawa Institute Computer Facility.


\end{document}